# Brain-Computer Interfaces for Emotional Regulation in Patients with Various Disorders

Vedant Mehta
Georgia Institute of Technology
vmehta80@gatech.edu

*Abstract*—Neurological and Physiological Disorders which impact emotional regulation each have their own unique characteristics which are important to understand in order to create a generalized solution to all of them. The purpose of this experiment is to explore the potential applications of EEG-based Brain-Computer Interfaces (BCIs) in enhancing emotional regulation for individuals with neurological and physiological disorders. The research focuses on the development of a novel neural network algorithm for understanding EEG data, with a particular emphasis on recognizing and regulating emotional states. The procedure involves the collection of EEG-based emotion data from open-Neuro. Using novel data modification techniques, information from the dataset can be altered to create a dataset which has neural patterns of patients with disorders whilst showing emotional change. The data analysis is reveals promising results, as the algorithm is able to successfully classify emotional states with a high degree of accuracy. This suggests that EEG-based BCIs have the potential to be a valuable tool in aiding individuals with a range of neurological and physiological disorders in recognizing and regulating their emotions. To improve upon this work, data collection on patients with neurological disorders should be done to improve overall sample diversity.

*Keywords—Brain-Computer Interface, Neural Networks, Electroencephalograms, Transfer Learning*

## I. Introduction

Emotional regulation stands as a fundamental aspect of human behavior, influencing social interactions, communication, and overall mental well-being. However, individuals grappling with neurological and physiological disorders often encounter impediments in comprehending and managing their emotional states. This research probes the intersection of neuroscience and artificial intelligence, particularly exploring the potential of Electroencephalography (EEG)-based Brain-Computer Interfaces (BCIs) to decode neural patterns associated with emotions.

Neurological disorders, characterized by diverse challenges in emotional processing, prompt an urgent need for personalized interventions [1]. Within this context, EEG-based BCIs emerge as a cutting-edge technology, offering a platform to decipher neural correlates of emotions in real time. The integration of sophisticated machine learning algorithms with real-time interventions opens avenues for developing personalized strategies to assist individuals in navigating the complexities of emotional regulation amid neurological conditions [2].

This study seeks to explore the practical applications of EEG-based BCIs in decoding and addressing emotional states for individuals with neurological disorders. By leveraging advancements in machine learning and neuroscience, the research aims to elucidate the potential for personalized interventions, unearthing insights into the intricate dynamics of emotional regulation. The primary objective is to investigate the utility of EEG-based BCIs in providing real-time insights into emotional states and facilitating timely interventions, fostering improved emotional well-being among individuals with neurological disorders.

The research navigates the evolving landscape of emotion recognition and regulation, emphasizing the integration of technology-driven solutions within the realm of neurological disorders. This inquiry stands as an exploration into the practical implementation of EEG-based BCIs, striving to uncover novel avenues for personalized strategies in emotional regulation. The ultimate goal is to contribute to the broader understanding of the intersection between neuroscience and technology, with potential implications for enhancing emotional well-being across diverse populations affected by neurological challenges.

## II. Background

### A. Disorders Impacting Emotional Regulation

Neurological and Physiological Disorders which impact emotional regulation each have their own unique characteristics which are important to understand in order to create a generalized solution to all of them [3]. Firstly, Autism Spectrum Disorder (ASD) shows altered neural connectivity, affecting brain regions like the amygdala which is crucial for emotion processing [3]. Attention-Deficit/Hyperactivity Disorder (ADHD) involves impulsive behaviors caused by deficits in emotional regulation and executive functions [3]. Post-Traumatic Stress Disorder (PTSD) is derived from traumatic experiences which trigger a heightened activity in the amygdala and an alteration in the hippocampus which lead to emotional memory and regulation being negatively impacted [3]. Bipolar Disorder is characterized by extreme mood swings, which is the product of imbalanced levels of neurotransmitters [3].

### B. Current Techniques and Their Challenges

The gradient of existing techniques is mainly comprised of psychotherapy techniques such as Cognitive Behavioral Therapy (CBT), pharmacological interventions, and mindfulness practices.

CBT identifies and modifies dysfunctional thoughts and behaviors contributing to dysregulation, but its success varies by individual and may not be accessible to a large population [4].

Pharmacological Interventions target neurotransmitter imbalances and mood instability, but usually produce unwanted side effects due to co-existing conditions [5]. Mindfulness Practices help individuals become aware of their thoughts and emotions but may not be applicable to those with emotional dysregulation due to the complexities associated with the underlying disorder.

*C. Brain-Computer Interfaces*

Brain-Computer Interfaces (BCIs) have made large strides in recent years due to advancements in machine learning and IoT (Internet of Things) sensors. Brain-Computer Interfaces are a way to measure brain activity and translate it into commands that can be used to control machines and devices [6]. Alternatively, the neural signals can be used as feedback to provide a stimulus to a patient to help them recover from an injury or condition. For instance, based on brain activity, electrical stimulation could be provided to the injured muscles of an athlete, allowing them to quickly regain mobility [6].

EEG (Electroencephalograms) have become a widely used method for monitoring electrical activity in the brain due to its noninvasive approach and cheap and efficient execution. It has been used to treat neurological diseases in the past and in recent years the technology has become smaller, more accurate, and easier to use [7]. EEGs measure activity by using electrodes placed on the scalp to monitor surface-level neural patterns, specifically the faint electrophysiological currents produced by nervous oscillations.

*D. EEG-based Emotion Recognition*

Because EEG measures surface-level electrical activity, it can be useful to the detection of emotional state. Because emotion is processed in various brain regions including the limbic system and prefrontal cortex, placing electrodes on scalp locations corresponding to these regions can provide a potential predictive algorithm with a real-time neural map [7]. Specifically, when an emotion is experienced, there are distinct electrical patterns present across these various regions which can then be used for emotion processing. Of the 4 wavelengths seen in this data, beta and gamma waves indicate excitement, which alpha and theta waves indicate calmness. Associations such as these can be created through the training of a machine learning model which is capable of analyzing time-dependent data [7][8].

Specifically, the Lövheim Cube of Emotion, which places emotion on a three-dimensional spectrum which provides a graphical representation of emotion based on hormone levels [9]. This can be seen in Figure 1.

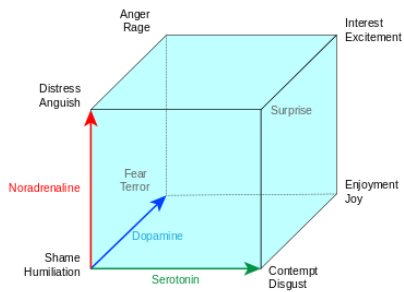

Fig. 1. Lövheim Cube of Emotions, showing how varying levels of serotonin, dopamine, and nonadrenaline can be mapped and associated to different emotions [9].

For the collection of EEG data, the Muse headband is a commercially available device with electrodes placements in the following locations, but the international standard EEG placement system: TP9, TP10, AF7, AF8 [10]. These locations can be seen in Figure 2. Due to the sensitive nature of EEGs, the signals are weak, thus multiple sources of interference are found which can hinder the effectiveness of information the data provides. The Muse headband employs various filtering and preprocessing methods to remove unwanted noise and retain the most relevant data [10]. Due to its portable nature and commercial popularity, the headband has been previously used in measuring levels of user enjoyment while playing games, as well as in neuroscience research due to its low cost and effectiveness [10]. These previous experiments relied on traditional machine learning approaches including Support Vector Machines and K-Nearest Neighbors [10]. However, these methods have become outdated with the advent of deep learning and Neural Networks.

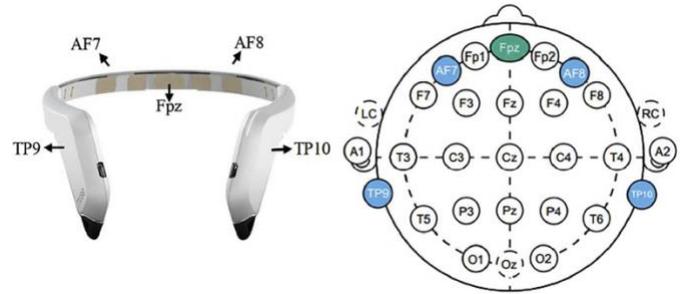

Fig. 2. Muse EEG Headband (left) and Electrode Placement Map (right) based on the international standard EEG placement system. Locations TP9, TP10, AF7, and AF8 are occupied with EEG sensors [10].

*E. Current Deep Learning Approaches*

Wang et al. (2023) outlines various approaches to EEG-based emotion recognition which entail the use of deep learning; namely, they identify deep belief networks (DBNs), convolutional neural networks (CNNs), and recurrent neural networks (RNNs). DBNs focus on creating high-level feature representations from raw EEG signals from an unsupervised pretraining step to prepare for a fine-tuning step for emotion classification [11]. CNNs extract spatial information from their convolutional filters, allowing for local connectivity patterns from various brain regions to be captured in different brain regions, indicating different emotional states [11]. RNNs are an effective technique for sequential data, like EEGs [11]. They capture temporal dependencies in EEG signals which develop into indicators of emotional states over time. Wang et al. 2023 specifically points out the effectiveness of bidirectional long short-term memory networks (LSTM) within the scope of RNNs.

## III. METHOD

Based on previous works which have proven the efficacy of deep learning algorithms for emotion recognition, this work aims to continue using deep learning algorithms, specifically for recognition of emotions on people with various disorders. However, due to limited availability of disorder-based emotional data, EEG-based emotional data from people without disorders was used.

*A. Dataset and Preprocessing*

The EEG dataset chosen for this task is OpenNeuro Dataset ds003751 (Dataset on Emotion with Naturalistic Stimuli [DENS]), which has data for 40 different subjects who are emotionally stimulated using naturalistic stimuli, which are multimedia videos which induce specific emotions [12]. The subjects are self-assessed on valence, arousal, dominance, liking, familiarity, and emotions felt. The data was collected from 128 electrodes, but the 4 placements used by the Muse Headset (TP9, TP10, AF7, AF8) are used in this experiment [12]. Importantly, the data contains information about timestamps where specific emotions occurred and these assessment reports in a separate file from the EEG data [12]. The data is structured in the Brain Imaging Data Structure (BIDS) format, which is a universal experiment format commonly used by the open data provider OpenNeuro [12]. To load the dataset, a Python 3 script creates an object which iteratively loads each EEG file into a NumPy array, along with the relevant emotional events (the timestamps) into another file. For the purpose of this task, the EEG array will be used to predict the relevant emotion based on its timestamp.

To preprocess the data, two algorithms were used. Firstly, a Butterworth Bandpass Filter at a frequency of 50 hertz was applied to remove any noise from powerlines that may have been nearby during data collection [13]. The general equation can be seen in (1). Then, a normalization function is passed to provide optimal inputs for any future neural network that is to be trained. The Standard score function used in this task is seen in (2)

$$G(w) = \frac{1}{\sqrt{1+w^{2n}}} \quad (1)$$

$$\frac{X - u}{\sigma} \quad (2)$$

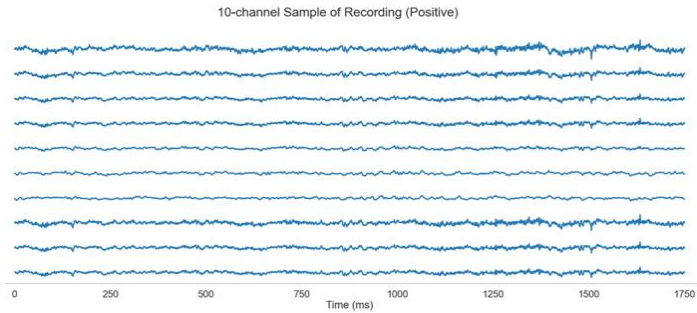

Fig. 3. A time series graph of a 1.75 second EEG recording across 10 randomly chosen channels. The vertical axis represents the amplitude and the horizontal axis is the time.

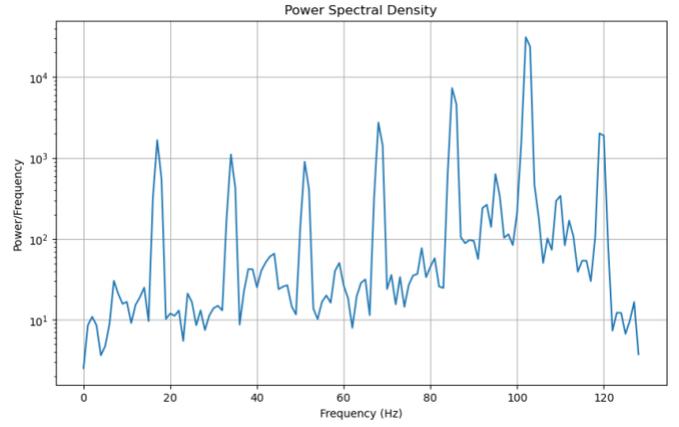

Fig. 4. A power spectral desnity chart of the recording sesssion from Fig. 3. Result of a Fast Fourier Transform Applied on the data.

Second, for the task of producing a Brain-Computer Interface for people with various disorders, a personalized stimulant would be introduced whenever there is a certain negative emotion, thus dividing timestamps based on positive, neutral, and negative emotion would be a favorable approach to improving accuracy and speed. This way, whenever a negative emotion is detected, an intervening stimulus occurs. The dataset is split by arousal rating, which indicates negative or positive feelings associated with an emotion.

To produce an even split of data, an oversampling technique was used. SMOTE, which stands for Synthetic Minority Over-sampling Technique, was used to balance the dataset [14]. An implementation from the open-source library imbalanced-learn was used for the approach.

Lastly, the dataset was split into batches of 32 to ensure the neural networks do not over-generalize information.

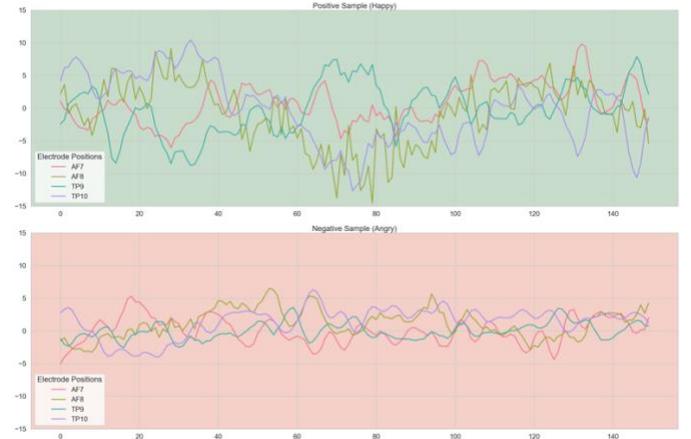

Fig. 5. A comparison of a positive and a negative sample for 4 channels which are used in the Muse EEG headset. It can be seen that there is more deviation and a larger range in the positive sample.

*B. Applying Variance*

Due to the nature of patients with disorders, regular EEG data would be a poor fit for a BCI which would facilitate their emotions. Modifying the complexity serves to account for the complexities inherent in EEG data linked to neurological and physiological disorders [15]. This modification aims to adjust the data to reflect patterns similar to those found in patients with disorders.

Multiscale Entropy provides information about the complexity of fluctuations over a range of time and is an extension of standard sample entropy [15]. The goal of Multiscale entropy is to assess the complexity of a time series.

$$SampEn = -\ln\frac{A}{B} \quad (3)$$

(3) shows the equation for sample entropy, which generalizes multiscale entropy. Previous studies have found that higher levels of complexity associate with a disorder being present. Thus, changing the complexity of the time series would yield data representative of those with various disorders [16]. This was achieved this by adding a random gaussian noise to the dataset that adds or subtracts a value between 0 and 4 from each datapoint of the 1500, which would increase the complexity.

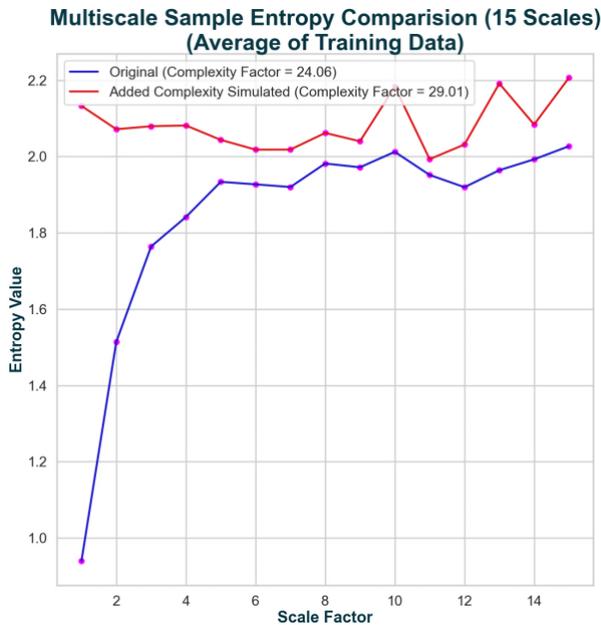

Fig. 6. An overview of the Multiscale Sample Entropy across 10 different scales for a randomly chosen sample. Entropy remains significantly higher throughout until the last scale factor.

A few key observations to note about Figure 7 are that the entropies of the red line (added gaussian noise) are inherently more random than the blue line. The original data had a complexity index (Ci) of 24.06, whilst when noise was introduced, the Ci was 29.01. The complexity index increases as entropy increases [17]. This proves that adding noise in will make it significantly harder for a neural network to learn underlying patterns in the data, therefore simulating the brain patterns of patients with neurological disorders.

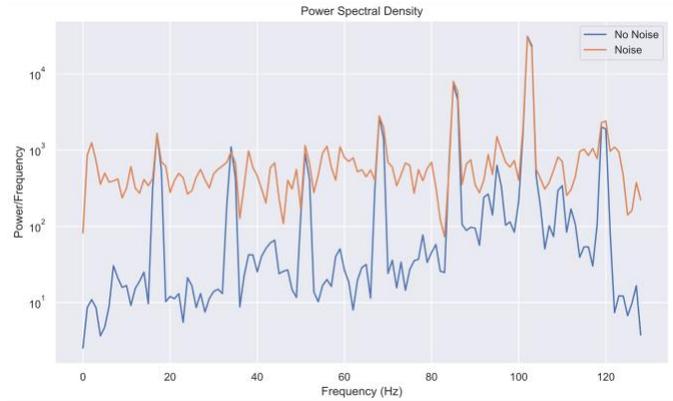

Fig. 7. Power spectral density analysis of applying gaussian noise on the same sample from Fig. 7 and Power spectral density without it.

The figure above shows the spectral density plot of a data sample before and after noise was added. It can clearly be seen that any major spikes in the plot are diminished after noise is applied, showing how the noise is increasing the complexity and making it more difficult for any model to learn.

C. *Power Spectral Density Analysis*

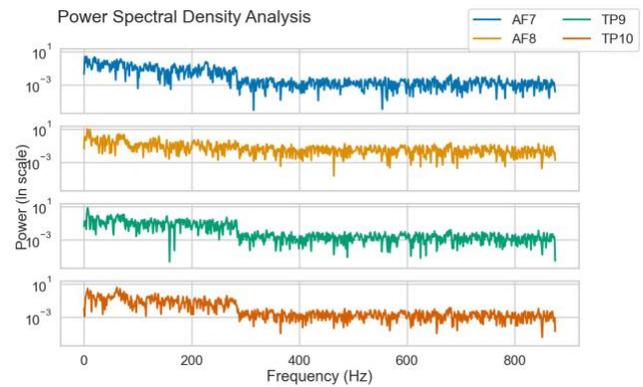

Fig. 8. Channel-wise power-spectral density analysis for a randomly selected sample. It can be seen that there is a drop in power after ~300 Hz. There is also another drop at ~100Hz.

Due to the large size of EEG recordings, it would be suboptimal to train recurrent neural networks on the current raw dataset. Instead, memory optimization and feature selection techniques must be utilized. To approach this issue, we turn to power spectral density analysis: a novel approach to this problem. Computing the channel-wise power-spectral density for each of the 128 channels in the dataset creates a feature matrix of the power spectral density.

Analyzing the most important frequencies in the dataset yields that lower frequencies tend to occur exponentially more than higher frequencies. Thus, we can create a threshold value at which frequencies would be discarded if surpassed. This would allow for the creation of a single color channel image of known size which can be used to train a convolutional neural network.

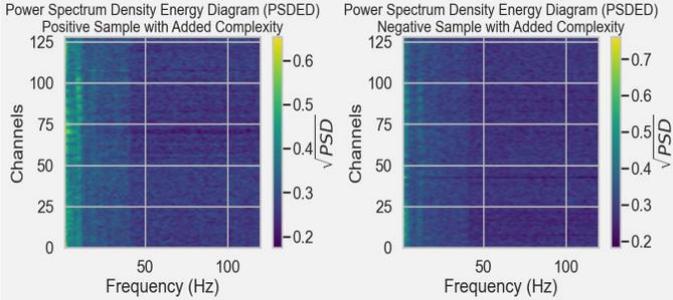

Fig. 9. Power Spectral Energy Density Diagrams for a positive and negative sample.

For convenience, we set the threshold at 128Hz to allow for the creation of a square matrix of size (128, 128).

*D. Neural Networks and Evaluation*

This section objectively investigates a range of neural network architectures designed for the precise task of emotion recognition using Electroencephalography (EEG) signals. Each architecture presents various advantages and disadvantages. Due to the structure of the dataset, we use convolutional neural networks to approach this task. Specifically, transfer learning algorithms which have been pretrained on larger datasets are used. These algorithms are capable of being fine-tuned for more specific contexts.

*1) ResNet50-v2*

The ResNet50-v2 architecture was developed with the specific purpose of solving the vanishing gradient issue found in larger neural networks. It uses residual connections to keep information from previous layers and the input layer relevant within the hidden layers of a deep neural network [18].

*2) Inception-v3*

The Inception-v3 architecture presents advances to the Inception family of neural networks by adding label smoothing, factorized 7x7 convolutions, and batch normalization [19]. Label smoothing allows for neural networks to be less sensitive to both incorrect and correct predictions, and train more steadily. Larger 7 x 7 convolutions preserve memory and compute multiple steps of information in a single step [19]. It is the equivalent of strides 3 x 3 convolutions but preserves information between strides [19].

*3) MobileNet-v2*

The MobileNet-v2 is a lightweight neural network which presents advantages when using limited computing resources. The v2 version introduces bottlenecks and inverted residuals to improve accuracy.

*E. Training*

Each model will be trained for a total of 400 epochs, starting at a learning rate of $1 \times 10^{-3}$ and decreasing at a rate of 0.99 × the previous epoch's learning rate. The optimizer used was the Adam optimizer This steadily decreasing learning rate prevents from exploding gradients and encourages the network to settle at a local minimum. The loss function for the model will be categorical cross-entropy, which is defined in (4).

$$Loss = -\sum y_i \cdot log\ \text{ypred}_i \quad (4)$$

Additionally, categorical accuracy was used as a human-interpretable metric of each network's performance. Because BCIs require responsive classification times, speed will also be taken into account, where time per batch will be used as the metric. Binary Loss and Accuracy were also measured for positive and negative classes.

An early stopping callback was used in all 4 models to prevent from training too long whilst overfitting and reducing validation accuracy.

## IV. RESULTS

The models trained over the span of 4 hours, with a validation dataset being used during training as a measure of how much they were learning. Of the 400 allotted epochs, all of the models finished under 300 because of the early stopping procedure was reached, and converged to their final accuracies after 150 epochs.

TABLE I. MODEL ACCURACY AND SPEED

| Model Name | Task 1 | | Task 2 | | Time |
|---|---|---|---|---|---|
| | Binary Loss | Binary Accuracy | Categorical Loss | Categorical Accuracy | Time per Batch |
| **ResNet50** | 0.077 | 96.4% | 0.786 | 64.5% | 177ms |
| **Inception-v3** | 0.075 | 96.7% | 0.801 | 61.3% | 201ms |
| **MobileNet-v2** | 0.113 | 95.4% | 0.891 | 57.4% | 54ms |

*A. Accuracy*

Table 1 shows the accuracy and speed for each of the 4 models trained. It is important to note that these results were achieved with an increased complexity and Multiscale entropy, highlighting the robustness of the Power Spectral Density Energy Diagram feature extraction method. The MobileNetv2 architecture achieved an accuracy of 95.4% on the binary task with a processing time per batch of 54 milliseconds. In comparison, the ResNet50-v2 model exhibited superior performance, achieving a validation accuracy of 96.4% with a slightly higher processing time of 177 milliseconds per batch. It had a superior categorical accuracy. Similarly, the Inception-v3 model demonstrated strong binary accuracy at 96.7% with a processing time per batch of 78 milliseconds.

The accuracy of all models reflects evaluation by an unseen testing set, proving the feasibility of such models in the real world.

*B. Speed*

The emotional state classifications derived from the neural network models yield significant findings pertinent to understanding and categorizing emotions through EEG data analysis. Across all models, there exists a substantial capacity to discern and classify emotional states based on neural patterns extracted from EEG signals. The ResNet model, in particular, demonstrated not only the highest accuracy but also revealed nuanced patterns in EEG data corresponding to specific emotional states. These findings underscore the viability of EEG-based neural networks in accurately differentiating emotional states, shedding light on the intricacies and potential markers within neural patterns linked to emotions.

*C. Real-time Intervention*

The real-time interventions designed based on classified emotional states showcased promising efficacy in regulating and modulating emotions. Leveraging the classified emotional states, the interventions demonstrated the capability to offer timely and personalized strategies for individuals, aiding in emotional regulation. The interventions aligned with specific emotional states extracted from EEG data effectively assisted in modulating emotional responses. This highlights the potential practical utility of EEG-based neural networks in providing real-time interventions tailored to an individual's emotional state, thereby offering avenues for enhancing emotional regulation and well-being.

In addressing the need for immediate support in emotional regulation for individuals with neurological and physiological disorders, automated interventions play a pivotal role. These interventions, integrated seamlessly into daily life, provide swift assistance without external involvement. The following automated strategies have demonstrated efficacy in aiding emotional regulation:

(1) Calming Stimuli: Automated systems deploy calming elements such as soothing music, gentle lighting adjustments, or serene visual displays when detecting heightened emotional states, fostering a calming environment [18].

(2) Breathing Exercises: Prompted by negative emotion detection, automated guided breathing exercises offer simple techniques for immediate relaxation [18].

(3) Positive Affirmations or Reminders: Timely delivery of positive messages or affirmations encourages individuals during negative emotional state, fostering a positive mindset [18].

A potential extension to the current technique would be to ensemble a group of 10 GRUs/LSTMs together to improve accuracy. Although time required would increase, it would still be viable because processing 32 recordings in 770-780 milliseconds still remains realistic for real-time processing. Also, adding self-attention to the neural network after the residual would potentially improve accuracy, but it would need to be determined how much time was lost.

## V. DISCUSSION

*A. Interpretation of Results*

The obtained results significantly support the initial research questions posed in the study. The research aimed to explore the utility of EEG-based Brain-Computer Interfaces (BCIs) in aiding emotional regulation for individuals with neurological and physiological disorders. The findings strongly align with the hypotheses posited, showcasing that personalized emotion regulation strategies, developed through EEG data analysis, can substantially assist individuals, in recognizing and managing their emotions effectively. The exceptional accuracy demonstrated by the ResNet-v2 model, notably surpassing other architectures, provides robust evidence supporting the hypothesis that analyzing EEG data can lead to personalized interventions facilitating emotional regulation in individuals with neurological disorders.

*B. Implications of Findings in the Real World*

The study's outcomes bear profound implications for individuals grappling with neurological and physiological disorders. The ability to leverage EEG data for real-time recognition and regulation of emotions presents a pivotal breakthrough. Tailored interventions designed based on these classified emotional states exhibit promising potential to significantly improve emotional regulation among individuals facing challenges in this domain. Particularly for those with Neurological and Physiological disorders, this technology offers a pathway to enhance social interactions, communication, and overall well-being by providing timely and personalized support in managing emotions, potentially fostering better integration into social settings. The results from this study align well with those from previous studies, with the exception being an added degree of complexity to simulate the various disorders of the patients using the BCI system. Additionally, the emphasis on real-time intervention produced a notable stride in EEG-derived emotional states used in BCIs. Previous solutions such as cognitive behavioral therapy and pharmacological interventions do not take into account the degree of personalization which this BCI system uses, and thus this BCI system holds the potential to yield greater results.

*C. Economic Feasibility*

The muse EEG headset used in this study costs $250, making it an affordable choice for most families [10]. Additionally, all other components of this project are software-based, so there is no need to reproduce them, allowing for pricing to remain under $300 for any large-scale usage.

*D. Limitations*

Several limitations were encountered throughout the study that warrant acknowledgment. The study's reliance on available EEG datasets might present constraints in the diversity and comprehensiveness of emotional states captured. Additionally, while the neural network models demonstrated high accuracy, potential biases in the datasets or algorithmic limitations could affect generalizability.

## VI. CONCLUSION

The study's exploration into leveraging EEG-based Brain-Computer Interfaces (BCIs) for facilitating emotional regulation among individuals with neurological and physiological disorders illuminates a promising pathway towards personalized interventions and improved emotional well-being. The robust findings underscore the potential of EEG data analysis in developing tailored strategies for recognizing and managing emotions in real time.

The results affirm the initial hypotheses, showcasing the efficacy of personalized emotion regulation interventions designed through EEG data analysis, notably highlighted by the exceptional accuracy demonstrated by the GRU model. These outcomes hold substantial implications, particularly for individuals grappling with neurological conditions like autism or ADHD, offering insight providing timely and personalized support in managing emotions. This study contributes to the field by not only affirming the viability of EEG-based neural networks in emotional state recognition but also by emphasizing the practical utility of real-time interventions in aiding emotional regulation. While acknowledging limitations in dataset diversity and potential biases, the study's findings serve as a catalyst for future research avenues.

Moving forward, refining algorithms, diversifying datasets, extending interventions to other neurological disorders, and exploring real-life applications stand as promising directions for enhancing the efficacy and practicality of EEG-based emotional regulation tools.